\documentclass[12pt,preprint]{aastex}
\usepackage{psfig}
\usepackage{lscape}

\def\xmm{{\sl XMM-Newton}}

\def\chandra{{\sl Chandra\ }}

\newcommand{\etal}{et al.~}

\shortauthors{Wang \etal}
\shorttitle{{\sl XMM-Newton} Spectral Analysis of Intermediate-Mass Black Hole Candidates}

\begin{document}

\title{{\sl XMM-Newton} Spectra of Intermediate-Mass Black Hole Candidates: Application of a Monte-Carlo Simulated Model}
\author{Q. Daniel Wang\altaffilmark{1}, Yangsen Yao\altaffilmark{1}, Wakako Fukui\altaffilmark{1}, ShuangNan Zhang\altaffilmark{2}, \& Rosa Williams\altaffilmark{1,3}}
\altaffiltext{1}{Department of Astronomy, University of Massachusetts, Amherst, MA 01003}
\altaffiltext{2}{Department of Physics, University of Alabama, Huntsville, AL 35899}
\altaffiltext{3}{Department of Astronomy, University of Illinois, Urbana, IL 61801}

\begin{abstract}
  We present a systematic spectral analysis of six ultraluminous X-ray sources  
  (NGC1313 X-1/X-2, IC342 X-1, HoIX X-1, NGC5408  X-1 and NGC3628  X-1) observed with 
  {\sl XMM-Newton} Observatory. These extra-nuclear X-ray sources in nearby late-type 
  galaxies have been considered as intermediate-mass black hole candidates. We have 
  performed Monte-Carlo simulations of Comptonized multi-color black-body accretion disks. 
  This unified and self-consistent spectral model assumes a spherically symmetric, thermal 
  corona around each disk and accounts for the radiation transfer in the Comptonization. 
  We find that the model provides satisfactory fits to the {\sl XMM-Newton} spectra of the 
  sources. The characteristic temperatures of the accretion disks ($T_{in}$), for example, 
  are in the range of $\sim 0.05-0.3$ keV, consistent with the intermediate-mass black hole 
  interpretation. We find that the black hole mass is typically about a few times 
  $10^3~\rm{M_\odot}$ and has an accretion rate $\sim 10^{-6} - 10^{-5} ~\rm{M_\odot~yr^{-1}}$. 
  For the spectra considered here, we find that the commonly used multi-color black-body 
  accretion disk model with an additive power law component, though not physical, provides 
  a good mathematical approximation to the Monte-Carlo simulated model. However, the latter 
  model provides additional constraints on the properties of the accretion systems, such as 
  the disk inclination angles and corona optical depths.

\end{abstract}


\section{Introduction}

Probably the most exciting recent development in the field of black hole 
(BH) study
is the discovery of numerous candidates for intermediate-mass BHs 
(IMBHs) with masses in the range of $\sim 10^2-10^5 M_\odot$ 
(see Miller \& Colbert 2003 
for a recent review).  The presence of
such BHs was first proposed to explain ultraluminous X-ray sources
(ULXs), defined as point-like 
extra-nuclear X-ray sources observed in nearby galaxies and 
with  inferred isotropic X-ray luminosities in excess of $10^{39}
{\rm~erg~s^{-1}}$, about an order of magnitude greater than the
Eddington limit of a solar mass object (e.g., 
Fabbiano 1989; Colbert \& Mushotzky 1999; Miller et al. 2003a,b; 
Strohmayer \& Mushotzky 2003). While unambiguous detections of 
individual IMBHs currently do not exist, there are observational hints 
from studies of microlensing events, globular clusters, and centers of 
nearby galaxies (van der Marel 2003 and references therein).
Although ULXs actually represent a heterogeneous population, a majority
of them are likely to be accreting BHs. The
controversy is centered on the X-ray emission mechanisms and on the masses of 
the BHs (e.g., Makishima et al. 2000; King et al. 2001; Begelman 2002). 

The IMBH interpretation, though probably the most straightforward and 
exciting, has serious difficulties (e.g., King et al. 2001;
Kubota et al. 2002). In addition
to their high X-ray luminosities, many ULXs show convex-shaped
spectra, especially in the energy band $\lesssim $ a few keV. Such spectra
are characteristic of the ``soft state'' of accreting BH binaries
and are often approximated by black-body-like models
such as the multi-color disk model (MCD;
{\em diskbb} in the {\em XSPEC} spectral analysis software package (Arnaud 1996; 
e.g., Makishima et al. 1986). In the MCD model, each annulus of the 
axis-symmetric optically-thick accretion is assumed to radiate as a blackbody
with a radius-dependent temperature. The characteristic temperature  $T_{in}$
of the innermost portion of the disk is $\propto (\dot{M}/M)^{1/4}$, where
$M$ and $\dot{M}$ are the BH mass and the accretion mass rate.
However, $T_{in}$ inferred from the model fit is almost always too high for the
required high mass, assuming the Eddington limit on $\dot{M}$. Equivalently, the 
inner disk radius $R_{in}$ is much smaller than the last stable orbit for
a non-spin BH. Even more disturbing is that the inferred 
value of $R_{in}$ is sometimes found to be time-variable, in contrast to the 
soft-state of confirmed BH X-ray binaries, where $R_{in}$ is approximately 
constant for each source.

Furthermore, it has been shown recently that many X-ray spectra of ULXs cannot 
be satisfactorily described by a single MCD model, especially in observations
with good counting statistics and with a broad energy coverage (e.g.,
Miller et al. 2003a,b). The usual practice is then to fit such a
spectrum with an additive combination of an MCD and a
power law (PL; e.g, Miller 2003a,b; Cropper \etal 2003). 
The requirement of this latter component, which often becomes 
important at energies $\gtrsim$ a few keV and may extend up 
to 200 keV as indicated by stellar mass BH systems, 
suggests that a high temperature 
electron cloud (corona) exists
around an accretion disk, producing inverse-Compton scattering of
the disk photons (e.g., Kubota et al. 2002; Page et al. 2003). A spectral fit with this 
additive, phenomenological model combination typically leads to an acceptable 
fit and a much lower (apparently more reasonable) disk temperature. 
Nevertheless, the model is over-simplified in 
the following two aspects:
 
First, the extension of the power-law component 
straight to the low-energy limit of the spectrum is nonphysical. 
Because the power-law component is assumed to mimic the effect of 
the inverse Compton scattering,
a low-energy cutoff as in the MCD component must be present 
in the Comptonized component (see also Page et al. 2003). Neglecting this
low-energy cutoff could mis-characterize the spectral shape 
of the MCD component and could lead to an artificially high absorption in spectral fitting.

Second, the additive combination of the MCD and the PL components
does not account for the radiative transfer process or the removal of photons 
from the MCD component to the Comptonized component. 
Furthermore, this process depends on both disk photon and 
corona electron energies. Therefore, one may not, in general,
directly take the MCD normalization derived 
from the spectral model fitting to infer the inner disk radius 
or the BH mass, as realized by some authors 
(e.g., Kubota, Makishima \& Ebisawa 2001). Otherwise, 
the inferred (but probably nonphysical) inner disk radius, 
for example, may appear to vary 
significantly when an accretion system changes from its 
hard state to its soft state or vice versa, as in 
XTE J2012+381 (Campana et al. 2002).

These oversimplifications in the MCD+PL model certainly obscure
the physical dependence of the Comptonization on the corona 
and disk properties, and could also seriously 
affect the inference of accretion disk parameters.

Yao \etal (2003, hereafter Paper I) have recently developed a Comptonized multi-color disk (CMCD)
model. They use Monte-Carlo simulations  to 
directly generate the Comptonized X-ray spectra, removing the above 
mentioned over-simplifications and avoiding the complications
of using the two~(unrelated)-component model and then trying to correct for
various radiative transfer effects. 
While the MCD model is still used to describe the accretion 
disk emission, the Comptonized radiation is no longer an independent component.
This self-consistent treatment thus provides a new tool to constrain the
physical properties of the corona and to study its relationship to the 
accretion disk. But most importantly, the  model
enables us to recover the same original disk flux in a spectral fit,
which is essential to a reliable mass estimate of the putative BH.

In the present work, we apply the CMCD model to the analysis of
{\sl XMM-Newton}  spectra of six ULXs which have been 
suggested as IMBH candidates. Our main objectives are to check whether or 
not the CMCD model provides an adequate spectral description 
of these sources and to see what potential new insights 
we may gain from such an application. The sources and data
are described in \S 2, whereas the implementation of the model 
for this application is discussed in \S 3, which also includes a summary of
various corrections required to infer BH masses from the 
present model fits.  We also test the PL, MCD, and MCD+PL 
models and compare them with the CMCD model. We present the results 
of our spectral fits in \S 4 and discuss the 
implications and conclusions in \S 5.

\section{Selected ULXs and \xmm\ Data}

For our initial application of the CMCD model to the IMBH candidates, we 
concentrate on {\sl XMM-Newton} X-ray observations of six previous known 
ULXs in nearby galaxies ($D \lesssim 10$ Mpc; Table~1). The 
moderate spatial resolution of
{\sl XMM-Newton} observations (e.g., FWHM $\sim 6^{\prime\prime}$ at 1 keV) 
is sufficient to isolate the emission from these individual ULXs in
the galaxies. Compared to similar \chandra\ observations, {\sl XMM-Newton} 
observations typically had substantially higher sensitivities and covered 
a broader energy range (0.2--15 keV), 
particularly important for constraining the Comptonization-related parameters.
If these ULXs are indeed accreting IMBHs, they should 
then have lower  $T_{in}$ values than those of stellar mass BH systems
(if the effect of BH spins is not important).
The corresponding spectral shift of the disk emission to lower energies
increases the importance of the Comptonized component in the 
{\sl XMM-Newton} energy
range. Therefore, {\sl XMM-Newton} data alone may allow us
to constrain simultaneously both the disk emission and 
the effect of the Comptonization.

Table~1 lists our selected sources with salient parameters of the host 
galaxies and the corresponding {\sl XMM-Newton} observations.
These six sources are located in five galaxies; each with exposure longer 
than 10 ksec for good counting statistics. The source positions
and their offsets from galactic nuclei are listed in Table~\ref{SourcePosi}.
All these ULXs are ``persistent'' sources and have been
studied previously based on the data from {\sl ASCA}, \chandra, and/or {\sl XMM-Newton}. 

We obtained the X-ray data from the {\sl XMM-Newton} Science Archive
and used the Standard Analysis System (SAS version 5.4.1, 2003) for 
data reduction, following the procedure described in the ABC guide 
for {\sl XMM-Newton} Data Analysis (version 1.3, 2002).
We checked light curves of individual observations and filtered out time
intervals with significant contamination from soft background flares. The 
final effective live-time for each dataset is included in 
Table~1. We utilized  imaging data from all three X-ray 
detectors: the European Photon Imaging Cameras (EPIC): MOS-1, MOS-2, and PN.
The {\sl XMM-Newton} observations of the four sources,
NGC1313 X-1/X-2, IC 342 X-1, and HoIX (HolmbergIX) X-1 (sometimes referred as 
M81 X-9) have been reported previously, although
not all of the data were used in individual studies. Observations for both
NGC5408 X-1 and NGC3628 X-1 are presented here for the first time. 
The following is a brief summary of key results from existing work:

\begin{itemize}
\item NGC1313 X-1/X-2: These two ULXs have been studied by Miller et al.
(2003a,b), based on the XMM-Newton MOS and PN
observations, separately.
From various acceptable additive two-component models, chiefly MCD+PL, 
they infer $T_{in} \sim 0.15$ keV for X-1 and 0.16 keV for X-2.  
These inferred low disk temperatures have been used as the key evidence for the consistency with 
the IMBH scenario. Several {\sl ROSAT} HRI observations of these two sources show
variability on the order of a factor $\sim 2$ \citep{col02}. 
X-1  has a 
radio counterpart with a luminosity of $\sim 10^{35} {\rm~erg~s^{-1}}$
\citep {col95}. An optical counterpart with R magnitude of 
21.6 to X-2 has been reported by \citet {zam03}. Both sources are associated 
with H$\alpha$ nebulae \citep {pak03}. 

\item IC342 X-1: This is one of the most intensively studied ULXs and
has shown strong variability over the years. For example, its luminosity in 0.5 - 10 keV 
band was 1.3$\times$10$^{40} {\rm~erg~s^{-1}}$ in September 1993 \citep {oka98} and 
decreased to 4.1$\times$10$^{39} {\rm~erg~s^{-1}}$ in February 2000 \citep {kubo02}.
The {\sl XMM-Newton} observation, as used in the present work, 
displays a further decrease in the source's luminosity to 
5.0 $\times$10$^{39}  {\rm~erg~s^{-1}}$.
Furthermore, the source is probably associated with an SNR \citep {rob03}. 

\item HoIX X-1: This source has been observed extensively by {\sl Einstein},
{\sl ROSAT}, {\sl BeppoSAX}, and {\sl ASCA} over 20 years. These observations 
show a strong time variability of the source \citep {par01}, including
apparent spectral state changes. Its highest luminosity 
reached $\sim 1 \times$10$^{40} {\rm~erg~s^{-1}}$ \citep {wang02} during 
an {\sl ASCA} observation in April 1999.  
The source is also associated with a giant shell-like H$\alpha$-emitting 
nebula \citep {wang02}.  \citet {mil03b} have presented an analysis of the \xmm\ PN 
data. The $T_{in}$ value inferred from an MCD+PL model fit is 0.21--0.26 keV,
again supporting the IMBH interpretation.

\item NGC5408 X-1: \citet {kaa03} have reported a {\sl Chandra}  observation taken in May 2002,
which indicates a luminosity of 1.1$\times$10$^{40} {\rm~erg~s^{-1}}$ 
in the 0.3--8 keV band. They have also proposed radio and optical counterparts for 
the source. 

\item NGC3628 X-1: A strong time variability has been observed 
from this source.
Its 0.1-2 keV flux was drastically reduced by a factor of $\gtrsim 27$ between 
December 1991 and May 1994 \citep {dah95}. A 52 ks {\sl Chandra}  observation in December
2000 showed a luminosity of 1.1$\times$10$^{40} {\rm~erg~s^{-1}}$
in the 0.3-8.0 keV band  
\citep{stri01}. No counterpart in other 
wavelengths has been reported for the source. 

\end{itemize}

We extracted the {\sl XMM-Newton} spectral data 
from a circular region with a radius in the 
range of 20$\arcsec$--30$\arcsec$ around each source, depending on the source position. 
For each source, the MOS-1 and MOS-2 data were combined. 
Each spectrum was grouped to contain a minimum 25 counts per bin.
The corresponding background spectrum was taken from a concentric 
annulus, removing any apparent sources  enclosed. 
A response matrix file and auxiliary response file were 
produced using the SAS tasks {\it rmfgen} and {\it arfgen}. The MOS 
and PN spectra are jointly fitted to tighten the constraints on spectral
parameters.

\section{Description of the CMCD Model}

The construction of the CMCD model has been detailed in Paper I,
including a discussion of various assumptions, comparisons with previous
works, and an application to the broad-band {\sl BeppoSAX} spectra of 
the stellar mass BH candidate XTE J2012+381 in our Galaxy.
This application successfully removes
the need for the varying inner disk radius and makes the specific predictions 
of both the size of the corona and the inclination angle of the accretion disk
as well as a more reliable estimate of both $T_{in}$ and $R_{in}$ (Paper I).

We have also applied the model to the spectral analysis of
the two persistent X-ray binaries, LMC~X-1 and LMC~X-3, which contain 
BH candidates (Yao \etal 2004, Paper II). Our derived foreground absorption column 
density ($N_H$) values, BH masses, and system inclinations are all consistent with 
those from the independent measurements based on optical and X-ray 
grating spectral data.
These tests demonstrate the applicability and predictive capability 
of the CMCD model in the study of accreting BH systems.

Here we briefly describe the implementation of the CMCD model 
for the study of ULXs.
For simplicity, we assume that the corona around the disk is
spherical and that the electron energy distribution in the corona
takes a thermal form, as in some previous works (e.g. 
Titarchuk 1994; Hua \& Titarchuk 1995; Poutanen \& Svensson 1996).
In Paper I, 
we also used the "slab"  geometry, which may be considered to be the opposite
extreme of the spherical corona shape. However, we note that 
these two different geometries do not make significant differences in
the fitted values of our most interested parameters ($T_{in}$ 
and the model normalization). The thermal assumption is another 
approximation. As discussed by \citet{coppi99}, a more realistic electron 
energy distribution might be a hybrid between a thermal plasma 
and a non-thermal high-energy tail. The thermal part would dominate the 
radiative transfer process in the low energy band, whereas the non-thermal part
would be important at high energies (beyond a few tens of
keV). Because the \xmm\ energy band used here is only upto $\sim 10$ keV,
our results are insensitive to the deviation from the assumed thermal form. 

\begin{figure}
\centerline{
\psfig{figure=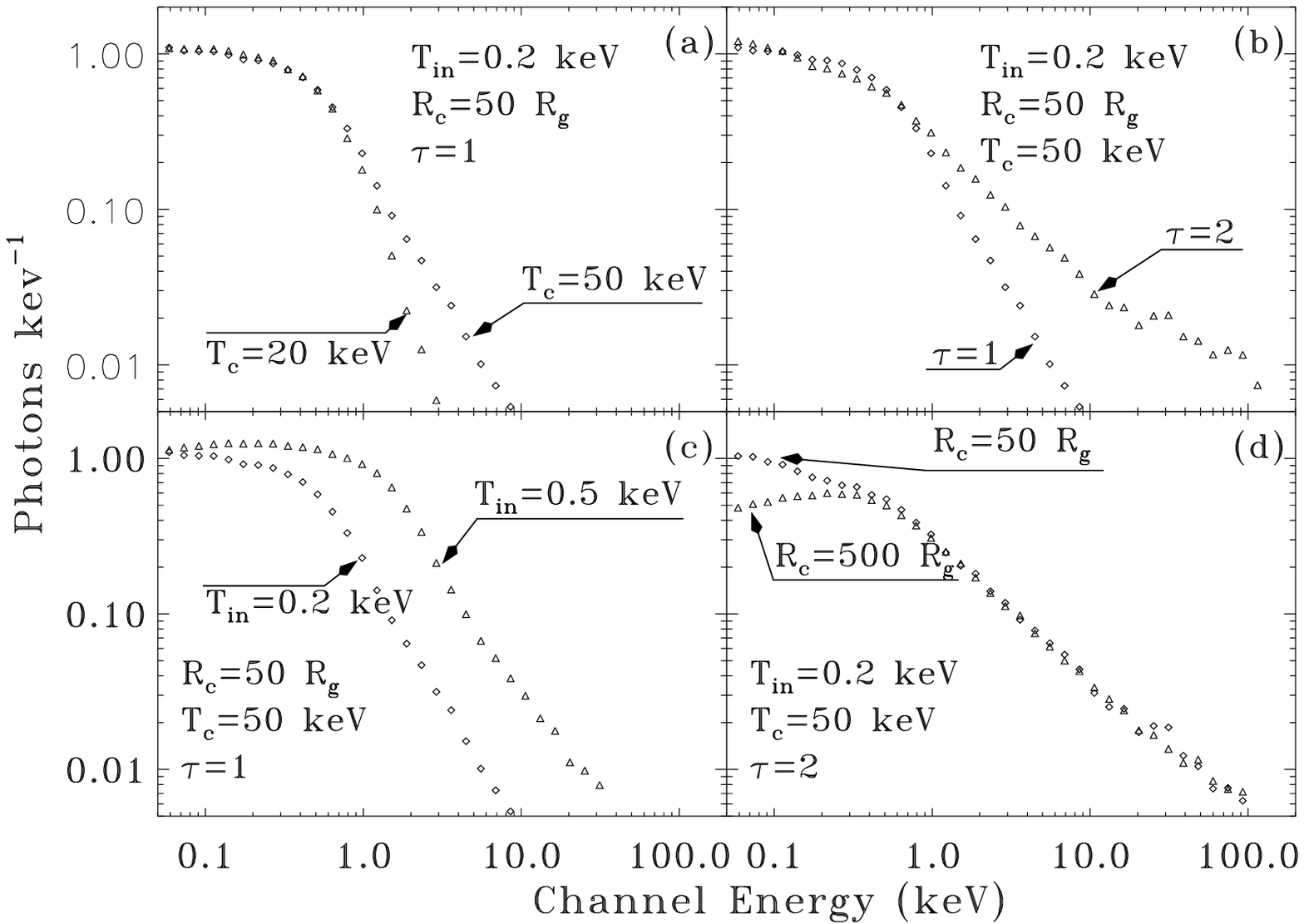}
}
\caption{\label{fig:demo}
Spectral dependencies on the electron temperature $T_c$ (a),
opacity of the corona $\tau$ (b), inner disk temperature
$T_{in}$ (c) and the corona size $R_c$ (d).
Definitions of the listed parameters are given in the text.
}
\end{figure}

The parameters of the CMCD model are the electron temperature ($T_c$), 
optical depth ($\tau$), and radius of the corona ($R_c$) as well as 
the inner disk temperature ($T_{in}$) and the disk 
inclination angle ($\theta$). Fig. \ref{fig:demo} illustrates the parameter
dependency of the inclination angle-averaged spectra.
The spectral dependency on the inclination angle is rather simple;
whereas the direct soft disk emission is proportional to cos~$\theta$, 
the Comptonized component is not affected.

Our implementation of the CMCD model is in a standard {\it XSPEC} table 
format\footnote{ftp://legacy.gsfc.nasa.gov/caldb/docs/memos/ogip\_92\_009/ogip\_
92\_009.ps}. This direct implementation avoids the unnecessary complications and
approximations in constructing an analytic expression (if possible)
and is convenient for any adjustments and changes in the model. The table 
contains a grid of spectra: $T_{in}$ and $\theta$ values are spaced by 
steps of 0.1 keV and $10^\circ$ linearly between the range 0.05-2 keV and 
between $22.7^\circ$--$79.2^\circ$, whereas $T_c$, $\tau$, and $R_c$ have 
4, 7, and 4 steps evenly spaced logarithmically between 10--100 keV, 
0--5, and 10--1000$R_g$ (where $R_g=GM/c^2$), respectively.

In a spectral fit, {\it XSPEC}  automatically interpolates between
the spectra in the table and employs a $\chi^2$ minimization 
algorithm. The normalization obtained from such a fit is similar to that 
of the MCD model
except without the cosine factor, since the disk projection effect 
has been taken into account in our simulation, i.e.
\[ K = \left(\frac{R_{in}/\mathrm{km}}{D/10\mathrm{kpc}}\right)^2, \]
where $D$ is the distance to the source and $R_{in}$  is the apparent 
radius with a peak disk temperature.

We estimate the mass of each putative BH as 
$M = {c^2R^\prime_{in} \over G\alpha}$, where
the inner disk radius, $R^\prime_{in}= f R_{in}$, is assumed to be the
same as the last stable circular orbit radius around the BH, $\alpha R_g$
($\alpha = 6$ or 1 for a non-spin or extreme spin BH). The factor 
$f= \eta (f_{col} f_{GR})^2 (\cos\theta/g_{GR})^{1/2}$ includes 
various corrections that have been dealt with in previous works
(Fig. 2):

\begin{itemize}
\item $f_{GR}$ and $g_{GR}$ relate the apparent and intrinsic radii
of the peak disk temperature \citep{zhang97}: $f_{GR}$ is due to the 
color temperature change caused by the gravitational red shift, and Doppler 
shift, whereas $g_{GR}$ is due to the integrated flux change caused by the 
gravitational focusing, time dilation and Doppler boostering. Both factors
depend on the BH spin and $\theta$ and account for General Relativity effects. 
We estimate the factors
from the quadratic interpolation of the tabulated 
values obtained by \citet{zhang97}.
$g_{GR}$ and $f_{GR}$ are in the ranges of 0.036 to 0.797 and of 0.355 to 
1.657, compared to $g_{GR}=\cos(\theta)$, and $f_{GR}=1$ in Newtonian case.

\item $f_{col}$ is the spectral hardening correction factor \citep{ebi84}.
Because of the high temperature at the inner disk region, which 
is responsible for the bulk of the emission, the 
inverse-Compton scattering at the surface of the disk becomes important
\citep{ross92}. The hardening of the spectrum effectively results in an 
underestimate of the inferred radius from the spectral fitting. This  
spectral hardening correction factor ($f_{col}=1.7$) 
has been calculated by \citet{shi95} by solving the disk structure
and radiative transfer self-consistently. The above corrections 
give the intrinsic radius for peak disk temperature.

\item $\eta$ is the ratio between the $R^\prime_{in}$ and the intrinsic 
radius for peak disk temperature. We use $\eta$=0.7 (non-spin) 
and 0.77 (extreme spin) derived by \citet{zhang97}, based on
a fully relativistic calculation. In Newtonian case, $\eta =1$.
\end{itemize}

\section{Results}

We summarize the results from our spectral fitting with 
the commonly-used MCD, PL, and MCD+PL models (Table~3) 
as well as the CMCD model (Table~\ref{results:cmcd}). 
The \xmm\ spectra of these sources are of the highest quality 
available for ULXs, which allows us to test whether or not they can be
characterized by these models. Table~3 shows that either PL or MCD alone can be rejected at a 
confidence greater than 4$\sigma$ for four out of the six sources. IC 342 X-1 
can be fitted reasonably well with either MCD or PL, although the latter is 
significantly better. The MCD model alone gives an acceptable fit to 
NGC3628 X-1, but the inferred $T_{in}$ 
value ($\sim 1.9 $ keV) is much
too high to be consistent with the IMBH interpretation. 
The additive MCD+PL combination is acceptable for all the sources, but it also
changes the same spectral parameters drastically. For example,
$T_{in}$ for IC 342 X-1 is reduced by a factor of $\sim 4$,
compared to the MCD fit. Let us now compare the results in Table~3
with those from the previous studies of the individual sources:

\begin{itemize}
\item NGC1313 X-1/X-2: Our results on these sources
are fully consistent with those 
from Miller et al. (2003a,b). Their results
are based on the same \xmm\ observations, but with MOS and PN data analyzed
independently, whereas ours are obtained from the joint-fits of the data. 

\item IC342 X-1:
Kong (2003) and Bauer et al. (2003) have shown that both PL and MCD models give satisfactory fits
to the same \xmm\ spectrum of this source. Our results are generally consistent with theirs.
Our $N_H$ value is slightly smaller than that obtained by 
Kong (2003; 5.14 $\times 10^{21}$  versus 6.0$\times 10^{21} {\rm~cm^{-2}}$), but is consistent
with that from Bauer et al. (2003) within the quoted statistical errors. 
Such small, though statistically significant, difference 
can presumably be due to various possible subtle differences in the data 
reduction and analysis (e.g., spectral extraction radius and binning). 

\item HoIX X-1:
Whereas individual spectra of this source 
from previous X-ray observations can be modeled satisfactorily 
with either PL or MCD (La Parola \etal 2001; Wang 2002), these models are not acceptable for the
{\sl XMM-Newton} data. Our MCD+PL results are consistent with those reported by \citet{mil03b}, 
who analyzed only the PN data. 
We find that the two separate observations give considerably
different spectral parameters, 
especially the PL index. Furthermore, the luminosity of the source during these \xmm\ 
observations is the highest known, about a factor of 2 greater than the previous record 
\citep {wang02}. 

\item NGC5408 X-1: 
The MCD+PL model gives a marginally acceptable fit to the \xmm\ spectrum. The same model
was shown to be satisfactory for the \chandra data (Kaaret et al. 2003). Our fitted spectral
parameters and the source luminosity are marginally consistent with those from Kaaret 
et al. (2003) within their respective uncertainties.  

\item NGC3628 X-1: Our obtained $T_{in}$ for the MCD fit is higher than 
the value from the \chandra data (Strickland et al. 2001), 1.87 versus 1.38 keV.
This change in $T_{in}$ could be due to the variability of the source. 
However, the higher $N_H$ value from the \chandra data may be due to the low-energy sensitivity 
degradation of the ACIS-S with time, which was apparently not corrected in the work (the 
correction software was only available recently).

\end{itemize}

Table~\ref{results:cmcd} shows that 
the CMCD model fits are satisfactory (or cannot be 
rejected at $\gtrsim 2\sigma$ confidence) to all the sources
(Fig.~\ref{fig:residual}).
Both $N_{\rm H}$ and $T_{in}$ are well constrained. 
In particular, the $T_{in}$ values are within 
a range of $\sim$ 0.1--0.3 keV, although the upper limit for IC342 X-1 
can reach $\sim 1.3$ keV.
For both $N_H$ and $T_{in}$ values, the CMCD and MCD+PL models are 
consistent with each other, within the statistical uncertainties.

\section{Discussions}

The satisfactory fits of the CMCD model to the {\sl XMM-Newton}
spectra of our selected ULXs suggest that they 
are consistent with the IMBH interpretation. In particular, the
model does not have the high $T_{in}$ problem as is faced by the 
MCD model. The problem is apparently caused by the neglect
of Comptonization in the model. Although this neglect
is statistically allowed when both the counting statistics and the energy
band coverage of an observed spectrum are poor, the fitted spectral parameters 
are far from being physical. We conclude that the MCD model alone should
not be used in the interpretation of ULXs as IMBHs.

We confirm that both $T_{in}$ and $N_H$ inferred from the MCD+PL model  
are reasonably accurate for the sources considered here. 
This apparent agreement between the 
CMCD and MCD+PL models suggests
that the latter model as a whole is mathematically a good representation of
the former model, at least for the IMBH candidates considered here.
This is rather surprising when one considers 
the over-simplifications in the MCD+PL model,
as discussed in \S 1. It appears that the nonphysical extrapolation of
the PL to the low energy nearly compensates the failure to include the 
radiation transfer loss of soft disk photons. 
However, this does not
mean that the MCD+PL model could be used to describe the 
Comptonized disk emission in general. 
The various nonphysical effects can cause problems for other
sources, especially those with higher
$T_{in}$ values ($\sim$ 1 keV; see the discussion in \S 3; e.g., LMC~X-1
and LMC~X-3; Paper II). 

In comparison,
the CMCD model provides more reliable measurements of the disk 
parameters as well as unique constraints on the physical 
properties of the coronae. In the following, we 
briefly discuss both the function of these new 
parameters and the physical reason for their different degrees 
of constraints:

\begin{itemize}
\item The opacity $\tau$ is relatively well constrained, which is the key
parameter that determines the total number of Comptonized photons (e.g., 
Fig.~\ref{fig:compare}). For example, IC 342 X-1 with 
the largest best-fit $\tau$ value appears to have the disk emission 
nearly completely Comptonized, explaining why the spectrum of the 
source can be characterized by a PL alone. Also for this source,
because the saturated Comptonization dominates over the thermal emission,
the constraints on $T_{in}$, $R_{in}$, and eventually on BH mass are very 
weak. 

\item The corona electron temperature 
$T_c$ is chiefly responsible for the overall energy extent of the 
Comptonized spectral component. Because $T_c$ is $\gtrsim 30$ keV for all
the sources, the high-energy turning-off of the component is well beyond the
\xmm\ band limit. Therefore, the data do not constrain the upper limit
of $T_c$.  The lower limit is determined because a minimum electron energy 
is needed to up-scatter soft disk photons to the high energies covered by
the spectra.

\item Whereas the nearly isotropic
Comptonized flux is barely affected by the disk inclination 
angle $\theta$, the observed strength of the soft disk component is 
proportional to cos($\theta$). In a  spectral fit, however, 
this difference in the disk inclination 
dependence may be partially compensated by a change in the $\tau$ value.
But if $\theta$ is large (for a nearly edge-on
disk),  its geometric effect cannot be canceled by adjusting other
parameters, which would also effectively alter the spectral shapes of both the 
disk and Comptonized components in a spectral fit
(e.g., Fig.~\ref{fig:compare}).
Consequently, we may constrain the upper limit to $\theta$. This
constraint, though not very tight, is important for the estimation of
the BH masses (\S 3). 

\item $R_c$ determines the effective corona radius, within which the disk 
emission is most affected by the Comptonization. Photons from
larger radii have relatively little chance
to be scattered and may contribute to the un-Comptonized disk component even 
if $\tau$ is large ($\geq 1$). But the amount of 
soft X-ray radiation from the disk also decreases with the increasing radius.
The combination of these two effects may thus place a constraint on $R_c$. 
\end{itemize}

Apparently, these parameters are correlated in a spectral fit.
This, together with the limited counting statistics and bandwidth of the data,
explains why the parameters are not tightly constrained. Nevertheless,
the results presented above demonstrate the potential of the CMCD 
model to shed new
insights into the physical properties of the accretion disk coronae,
in addition to a more reliable mass estimate of  the putative BHs.

Table~\ref{results:flux} includes our estimated BH masses,
assuming no spin and the best-fit $\theta$ values. The typical BH mass 
is in the range of $\sim 10^3-10^4$ M$_\odot$, although the
upper limit for IC~342~X-1 is slightly higher.
If a BH spins rapidly, the inferred BH mass could be several times higher
 than the value quoted in the table (depending on $\theta$; Fig. 2). 

Assuming that the bolometric luminosity (estimated in
the 0.05--100 keV range) $L_{bol} = 0.1  \dot{M}c^2$, we further 
estimate the accretion rate $\dot{M}$ for each source
(Table 5), which is in the range of 
1--10 $\times 10^{-6}$ M${\rm_{\odot}yr^{-1}}$.

The present work represents, at most, an incremental step in developing
a fully self-consistent model for accreting BH systems. The CMCD model
used here deals only with the Comptonization by static disk coronae. To study
the dynamics, one needs to understand the 
formation and evolution of the coronae as well as the physics of 
the accretion disks. We also have not considered other proposed 
scenarios that may explain some of the ULXs; e.g., the anisotropic 
emission of the 
radiation \citep{king01}, the relativistic motion of the 
X-ray-emitting plasmas \citep{fab01}, and the possible 
super-Eddington emission (Begelman 2002).
Spectral models for such scenarios, yet to
be developed, need to account for the apparent presence of the soft
thermal component, in addition to the power law, for these sources
except in the case of IC 342~X-1. The bottom line here is that the \xmm\
spectra are consistent with the IMBH interpretation of the sources.

\acknowledgments
We thank Xiaoling Zhang for useful discussions and the referee for
many thoughtful comments that helped in improving the presentation of the paper.
This work is supported by the NASA LTSA grant NAG5-8935.



\clearpage

\clearpage
\begin{figure}
\centerline{
  \psfig{figure=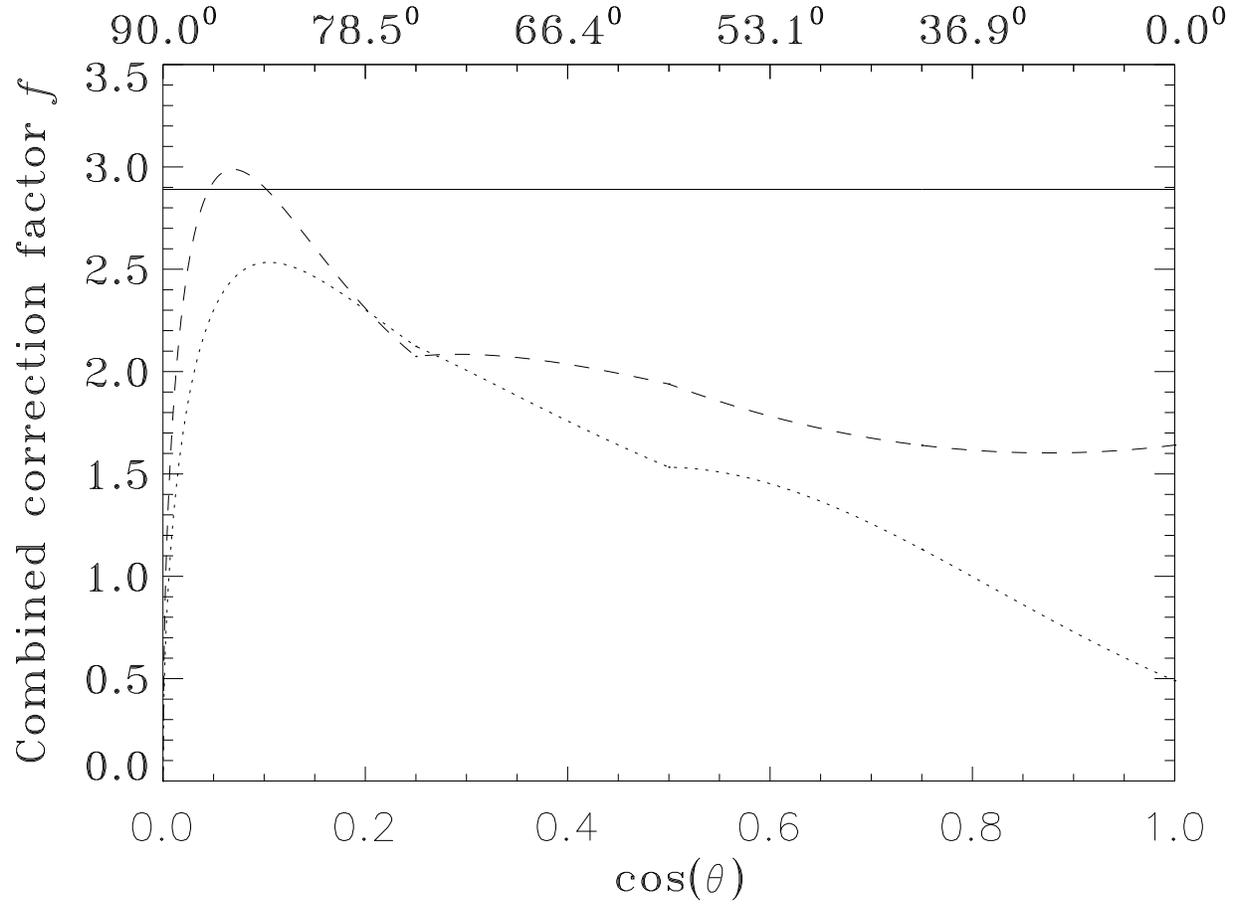}
}
\label{fig:CorrectionFactor}
\caption{
The combined $R_{in}$ correction factor $f$ vs. the disk inclination angle $\theta$ 
(the scale at the top), or cos($\theta$) (at the bottom). {\it Solid line}: Newtonian case; 
{\it dashed line}: non-spin case; and {\it dotted line}: extreme spin case. The figure is 
based on Table~1 of \citet{zhang97}.}
\end{figure}

\clearpage
\begin{figure}
\centerline{
  \psfig{figure=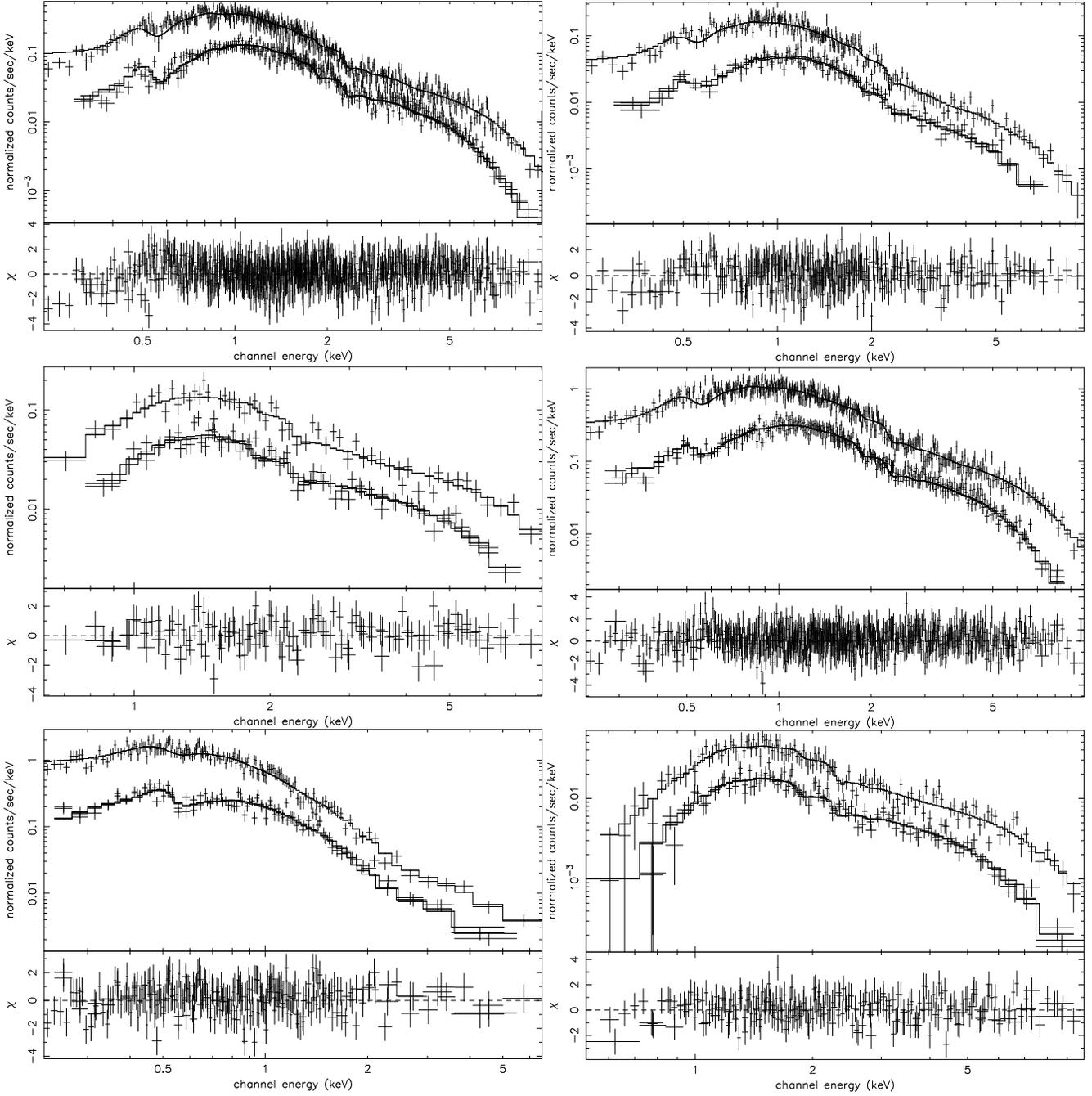,height=7truein,clip=}
}
\caption{\label{fig:residual}
 Model fits to the {\sl XMM-Newton} spectra of ULXs:
NGC1313 X-1 (top left), NGC1313 X-2 (top right), IC342 X-1 
(middle left), HoIX X-1 (OBS 1 only for clarity; middle right), NGC5408 X-1 (bottom left), and  
NGC3628 X-1 (bottom right). The solid line shows the fit of the CMCD model. For each source, 
the model is jointly applied to the PN spectrum (which always has a higher 
flux) as well as to the MOS 1 \& 2 spectra.}
\end{figure}

\clearpage
\begin{figure}
\centerline{
\psfig{figure=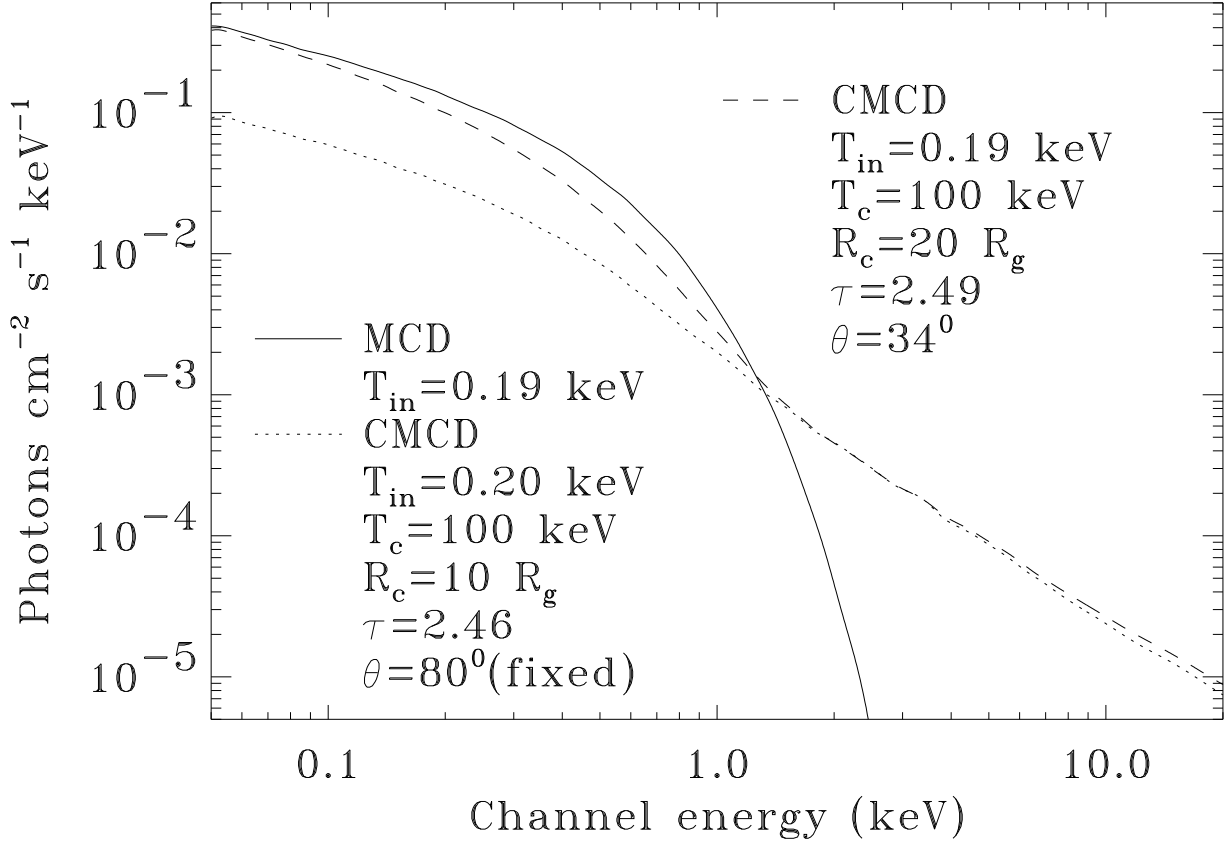}
}
\caption{Illustration of the spectral dependence on the corona opacity $\tau$ and disk inclination $\theta$
in the MCD model of HoIX~X-1: the absorption-corrected best-fit CMCD model ({\sl dashed line}); 
the corresponding MCD model before Comptonization ({\sl solid line});
the absorption-corrected best-fit CMCD model with inclination fixed at $\theta$ = 80$^0$ 
({\sl dotted line}).}
\label{fig:compare}
\end{figure}

\clearpage
\begin{deluxetable}{lcccccccccc} 
\tabletypesize{\scriptsize}
\tablecaption{Selected ULXs and {\sl XMM-Newton} Observations}
\tablewidth{0pt}
\tablehead{
\colhead{Galaxy} & 
\colhead{Hubble}    &
\colhead{$D$}&  
\colhead{N$_{\rm HI}$} & 
\colhead{OBS} &
\colhead{OBS Date} &
\colhead{MOS1} &
\colhead{MOS2} &
\colhead {PN} &
\colhead{Filter}\\
\colhead {}&
\colhead{Type} &
\colhead {(Mpc)}  &
\colhead {(10$^{21}$cm$^{-2}$)} &
\colhead{ID} &
\colhead{(mm/dd/yy)} &
\colhead {(ksec)}&
\colhead {(ksec)}&
\colhead {(ksec)}&
\colhead {}
}
\startdata
NGC1313 X-1/X-2 &  SB(s)d& 3.7& 0.40  & 0106860101 & 10/17/00 &28.9/29.3& 28.9/29.2 & 28.4/31.6 & medium \\
IC342 X-1               & Scd    & 3.3& 3.03    &0093640901 & 02/11/01  & 9.5/9.8 &  9.5/9.8 & 5.0/5.9 & medium \\
HoIX X-1                &   Im   & 3.6&0.40     &0112521001 & 04/10/02 & 10.0/10.1& 10.0/10.1 &  6.8/7.8 &  thin1\\
  (or M81 X-9)         &        &    &                 &0112521101 & 04/16/02 & 7.9/10.8 &7.9/10.8 & 7.4/8.5 & thin1     \\
NGC5408 X-1        & IB(s)m & 4.8& 0.57  & 0112290601 & 08/08/01 & 5.9/7.5 & 5.9/7.5 & 4.2/5.0 & thin1 \\
NGC3628 X-1        & Sbc    &10.0& 0.22   & 0110980101 & 11/27/00 & 42.2/54.5 & 42.3/54.6 & 32.2/50.7 & thin1  
\enddata
\tablecomments{$D$ is the galaxy distance: NGC1313 \citep {tul88}, IC342
\citep {saha02}, HoIX \citep {free94}, NGC5408
\citep {kar02}, and NGC3628 \citep {soi87}, whereas the
$N_{\rm HI}$ values are all from the Galactic HI survey by \citet {dickey90}.
Exposure time of MOS1, MOS2, and PN are listed as
the cleaned time interval used for our analysis/original exposure time.}
\label{SelectedSources}
\end{deluxetable}

\clearpage
\begin{deluxetable}{lcccc}
\tablecaption{Positions of the Selected ULXs}
\tablewidth{0pt}
\tablehead{
\colhead{ULXs} & 
\colhead{R.A.}    &
\colhead{Dec.} &
\colhead{$\delta$}    \\
\colhead{} & 
\colhead{(J2000)} &
\colhead {(J2000)} &
\colhead {($^{\prime}$)}     
}
\startdata
NGC1313 X-1 & 3:18:20.21 & -66:29:10.7 & 0.83\\
NGC1313 X-2 & 3:18:22.62 & -66:36:05.9 & 6.2\\
IC342 X-1 & 3:45:55.46 & 68:04:54.2 & 5.0 \\
HoIX X-1 & 9:57:53.50 & 69:03:47.8 & 2.2\\
NGC5408 X-1 & 14:03:19.62 & -41:23:00.2 & 0.42\\
NGC3628 X-1 & 11:20:16.23 & 13:35:15.0 & 0.12
\enddata
\tablecomments{The X-ray source positions are from the 
{\sl XMM-Newton} observations, whereas $\delta$ is the projected offset
of each source position from the respective galactic center obtained from NED.}
\label{SourcePosi}
\end{deluxetable}

\clearpage
\begin{table} 
\tabletypesize{\scriptsize}
\begin{center}
\caption{Results from spectral fits with 
the PL, MCD, and PL+MCD models}
\setlength{\evensidemargin}{1in}
\begin{tabular}{lcccccccccc}
\hline
\hline
Source & Model & $N_H$ ($10^{21}{\rm~cm^{-2}}$) & $\Gamma_p$ or $T_{in}$(keV) & $ K_{PL} or K_{MCD}$ & $\chi^2$/dof\\
\hline
NGC1313 X-1 &PL&2.11 (2.03--2.19) &1.95 (1.92--1.98)& 6.3$\times 10^{-4}$ & 1222/783 \\
            &MCD  &0.49 (0.44--0.53)        & 1.39 (1.36--1.43)   & 0.031 & 2217/783 \\
            &PL+  &4.43(4.00--4.86) & 1.81(1.78--1.85)   & 5.51$\times 10^{-4}$& 851/781 \\
            & MCD &                 &0.16(0.15--0.18)    & 628\\
NGC1313 X-2 &PL&2.82 (2.68--2.97) & 2.51 (2.44--2.57)  & 4.4$\times 10^{-4}$&474/352 \\
            &MCD  &0.78 (0.70--0.87)& 0.85 (0.81--0.88) & 0.092 & 858/352 \\
            &PL+ & 3.72 (3.14--4.41) & 2.23(2.14--2.36) & 3.3$\times 10^{-4}$  & 386/350 \\
            &MCD&                          & 0.18(0.16--0.20) & 99 \\            
IC342 X-1   &PL  &5.13 (4.56--5.75) & 1.66 (1.57--1.76)  & 5.4$\times 10^{-4}$ & 113/129 \\
            &MCD  &2.76 (2.39--3.16) & 1.91 (1.77--2.08) & 0.011 & 146/129 \\
            &PL+  &5.14 (3.82--9.19)   & 1.54 (0.78--1.76)  & 4.5$\times 10^{-4}$  & 112/127 \\
            &MCD  &                          &0.50 (0.10--0.93)   & 0.23 \\
HoIX X-1  obs.1 &PL& 1.81(1.73--1.89)& 1.86(1.83--1.89)&17$\times 10^{-4}$&852/707 \\
            &MCD  &  0.37(0.32--0.41) & 1.46(1.42--1.50) & 0.08  &  1593/707  \\
            &PL+  & 2.90(2.55--3.27) & 1.72(1.66--1.74) & 14$\times 10^{-4}$ & 706/705  \\
            &MCD  &                 & 0.20(0.18--2.23) & 220\\
HoIX X-1 obs.2 & PL& 2.14(2.07--2.22) &1.94(1.91--1.97)&22$\times 10^{-4}$ & 937/ 769 \\
            &MCD & 0.57(0.52--0.61) & 1.38(1.35--1.42) & 0.11 &1764/769  \\
            &PL+ & 3.49(3.08--3.94) &1.86(1.82--1.91) & 20$\times 10^{-4}$ & 810/767 \\
            &MCD &                        &0.17(0.16--0.19) & 689 \\
NGC5408 X-1 &PL  &1.96 (1.80--2.15) & 3.79 (3.67--3.93) & 13$\times 10^{-4}$& 393/260  \\
            &MCD  &0.40 (0.32--0.49) & 0.30 (0.29--0.31)       & 21                           &640/260 \\
            &PL+  &1.34(1.15--1.65) & 2.56(2.36--2.81)   & 4.7$\times 10^{-4}$ & 283/258 \\
            &MCD  &                 &0.18(0.16--0.19)    & 285\\
NGC3628 X-1 &PL  &6.77 (6.26--7.32) & 1.78 (1.71--1.86) &1.6$\times 10^{-4}$& 298/246 \\
            &MCD  & 3.84 (3.52--4.18) & 1.87 (1.77--1.99) & 0.003  & 398/246 \\
            &PL+ & 9.53(7.54--11.9) & 1.74(1.58--1.89) & 1.6$\times 10^{-4}$ & 285/244 \\
            &MCD &                        & 0.23(0.18--0.36) & 15\\
\hline
\end{tabular}
\tablecomments{$\Gamma_p$ is the photon index of the PL model.
The normalization of the PL model, $K_{PL}$, is defined as ${\rm photons~keV^{-1}~cm^{-2}~s^{-1}}$ at 1 keV, whereas the normalization of the MCD model, $K_{MCD}$, is defined as $(R_{in}/{\rm km})^{2}{\cos}\theta/(D/10~{\rm kpc})^2$. The uncertainty ranges of the parameters are all at the 90\% confidence.  
}
\end{center}
\end{table}

\clearpage
\begin{rotate}
\begin{table} 
\caption{Results from spectral fits with the CMCD model}
\footnotesize
\begin{tabular}{lcccccccc}
\hline
\hline
Source & $N_H(10^{21}~\mathrm{cm^{-2}}$) & $T_{in}$(keV) & $T_c$(keV) & $R_c(R_g)$ & $\tau$ & $\theta$ (deg) & $K (10^2)$ & $\chi^2/dof$ \\
\hline

NGC~1313~X-1    &4.11(3.98,4.27)         &0.199(0.159,0.201)     &100(95,--)     &11(--,15)      &2.7(2.4,3.1)        &23(--,34)      &3.4(3.1,6.5)   &860/779\\
NGC~1313~X-2    &3.25(3.09,3.73)         &0.19(0.12,0.20)        &99(68,--)      &49(24,84)      &1.0(0.87,1.4)       &23(--,58)      &1.3(0.89,3.1)    &381/348\\
IC342~X-1       &5.23(4.13,6.53)         &0.32(0.05,1.27)        &49(28,--)      &20(--,--)      &5.0(2.3,--)         &79(--,--)      &0.57(0.01,49)     &111/125\\
HoIX~X-1~obs.1  &3.35(3.12,3.57)         &0.19(0.12,0.20)        &100(76,--)     &19(10,28)      &2.5(2.2,3.1)        &34(--,59)      &8.8(5.5,35)  &710/703\\
HoIX~X-1~obs.2  &3.31(3.08,3.67)         &0.18(0.10,0.20)        &100(69,--)     &22(--,41)      &1.9(1.7,3.4)        &50(--,62)      &12(6.6,52) &816/765\\
NGC5408~X-1     &1.31(1.19,1.39)         &0.13(0.11,0.20)        &98(66,--)      &10(--,17)      &1.0(0.81,1.4)       &25(--,53)      &9.5(2.1,27)  &279/256\\
NGC3628~X-1     &9.79(7.97,11.85)        &0.21(0.12,0.31)        &100(33,--)     &22(--,42)      &2.6(2.1,--)         &23(--,73)      &0.49(0.1,4.7)     &283/242\\

\hline
\end{tabular}
\tablecomments{The upper limit of our current table model for $T_c$ is
100 keV. The symbol '--' indicates that the limit is not  
constrained. }
\label{results:cmcd}
\end{table}
\end{rotate}

\clearpage
\begin{table} 
\begin{center}
\caption{Inferred Parameters from the CMCD Model}
\label{results:flux}
\begin{tabular}{lccccccccccccc}
\hline
\hline
Source &$f_{2-10}$/$f_{0.2-10}$ &$L_{2-10}$/$L_{0.2-10}$ & log($R^\prime_{in}$/km) & log(M$_{\rm BH}$/M$_\odot$) &  ${\rm\dot{M}/\dot{M}_6}$\\
\hline
NGC1313 X-1      &  1.9/9.1 &  3.1/14.8 & 4.04(4.02,4.20) & 3.09(3.08,3.25)&  4.9\\
NGC1313 X-2      &  0.6/2.8 &  1.0/4.6  & 3.83(3.76,4.06) & 2.88(2.81,3.11)&  1.2\\
IC342 X-1        &  2.4/4.1 &  3.1/5.4  & 3.77(3.48,5.41) & 2.82(2.53,4.46)&  3.1\\
HoIX X-1 (OBS 1) &  5.7/19.5&  8.8/30.3 & 4.23(4.14,4.63) & 3.29(3.20,3.68)& 11.3\\
HoIX X-1 (OBS 2) &  6.3/20.8&  9.8/32.2 & 4.33(4.22,4.76) & 3.38(3.28,3.81)& 10.2\\
NGC5408 X-1      &  0.6/7.1 &  1.6/19.6 & 4.38(4.16,4.66) & 3.43(3.21,3.71)&  5.0\\
NGC3628 X-1      &  0.6/2.1 &  7.3/25.3 & 4.05(3.83,4.78) & 3.10(2.88,3.83)&  9.7\\
\hline
\end{tabular}
\tablecomments{The fluxes $f_{\rm 2-10}$ and $f_{\rm 0.2-10}$ have been corrected 
for the absorption and are in units of $10^{-12} {\rm~erg~cm^{-2}~s^{-1}}$ and 
are in the energy range 2--10 and 0.2--10 keV, respectively. 
The luminosities $L_{\rm 2-10}$ and $L_{\rm 0.2-10}$ are in units of 
$10^{39}{\rm~erg~s^{-1}}$ and are calculated from the fluxes and the 
distances in Table~1.
The {$\rm \dot{M}_6$} is 10$^{-6}$ {$\rm M_{\odot}yr^{-1}$}.
Assuming no spin and $L_{bol}=0.1 \dot{M} c^2$. See text for detail.
}
\end{center}
\end{table}

\end{document}